\documentclass[aps,prl,twocolumn,floatfix,10pt]{revtex4-1}
\usepackage{graphicx,bm}
\usepackage{times}
\usepackage{amsmath,amssymb}
\usepackage[usenames,dvipsnames]{xcolor}

\begin{document}

\title{Absence of diffusion and fractal geometry in the Holstein model at high temperature}

\author{Chen-Yen Lai}
\author{S. A. Trugman}
\affiliation{Theoretical Division, Los Alamos National Laboratory, Los Alamos, New Mexico 87545, USA}
\affiliation{Center for Integrated Nanotechnologies, Los Alamos National Laboratory, Los Alamos, New Mexico 87545, USA}

\date{\today}

\begin{abstract}
We investigate the dynamics of an electron coupled to dispersionless optical phonons in the Holstein model, at high temperatures.
The dynamics is conventionally believed to be diffusive, as the electron is repeatedly scattered by optical phonons.
In a semiclassical approximation, however, the motion is not diffusive.
In one dimension, the electron moves in a constant direction and does not turn around.
In two dimensions, the electron follows and then continues to retrace a fractal trajectory.
Aspects of these nonstandard dynamics are retained in more accurate calculations, including a fully quantum calculation of the electron and phonon dynamics.
\end{abstract}


\maketitle



In band theory, non-interacting electrons can carry current forever.
If interactions with impurities or phonons are included, in a simple picture,
after a collision, an electron loses memory of its previous state,
and the motion is diffusive.
Situations in which this picture fails have deepened our understanding of condensed matter physics.
It is not entirely correct to assert that successive electron-phonon scattering events are uncorrelated.
An electron that emits a phonon can coherently reabsorb the same phonon, forming a polaron quasiparticle,
which (at zero temperature) can again carry current forever, but with a different effective mass.
With respect to impurity scattering, under certain conditions diffusion does not occur (Anderson localization).
We consider the Holstein model of electron-phonon interactions.
We will argue that at high temperatures, in this case as well, the diffusive picture of electron transport can be misleading.

One of the most fundamental problems in condensed matter physics is how electron-phonon coupling leads to electrical resistivity in crystals.
We would like to reconsider the transport
of electron-phonon coupled systems at high temperatures.
A simple description of electron-phonon coupling is given by the Holstein model~\cite{holstein1959},
in which dispersionless optical phonons couple to the electron density.
The Holstein Hamiltonian is
\begin{eqnarray}\label{eq:holstein}
  H\!=\!-\bar{t}\sum_{\langle ij \rangle} (c^\dagger_ic_j \!+\! h.c) \!+\! \Omega \! \sum_j a^\dagger_j a_j \!-\! g \! \sum_j n_j(a_j \!+\! a^\dagger _j)\;,
\end{eqnarray}
where $c^\dagger_j$ creates an electron and $a^\dagger_j$ creates a phonon on site $j$, and the number operator is
$n_j\!=\!c^\dagger_jc_j$.
We consider the dynamics of a single electron.

What is the (DC) conductivity of the Holstein model?
With quantum phonons, the polaron quasiparticle is the exact ground state at any momentum k
(or at least at momenta near zero in higher dimensions).
These states carry current forever, so the conductivity (or mobility) is infinite at zero temperature.
At low temperatures, there are in addition thermally excited phonons that are exponentially dilute.
The polaron is expected to scatter from these phonons, leading to an exponentially large conductivity,
$\sigma \!\sim\! e ^ { \Omega / T }$ as $ T \!\rightarrow\! 0 $.
( $\hbar$ and $k_B$ have been set to 1.)
What about high temperatures, $T > \Omega$?
A plausible scenario is that from Boltzmann transport theory, the electron has diffusive behavior from scattering off thermally excited phonons,
leading to a finite conductivity from the Einstein relation~\cite{madelung1978introduction,mishchenko2015}.
We will argue below that this scenario is misleading in certain high temperature regimes.

\begin{figure}[t]
  \begin{center}
    \includegraphics[width=\columnwidth]{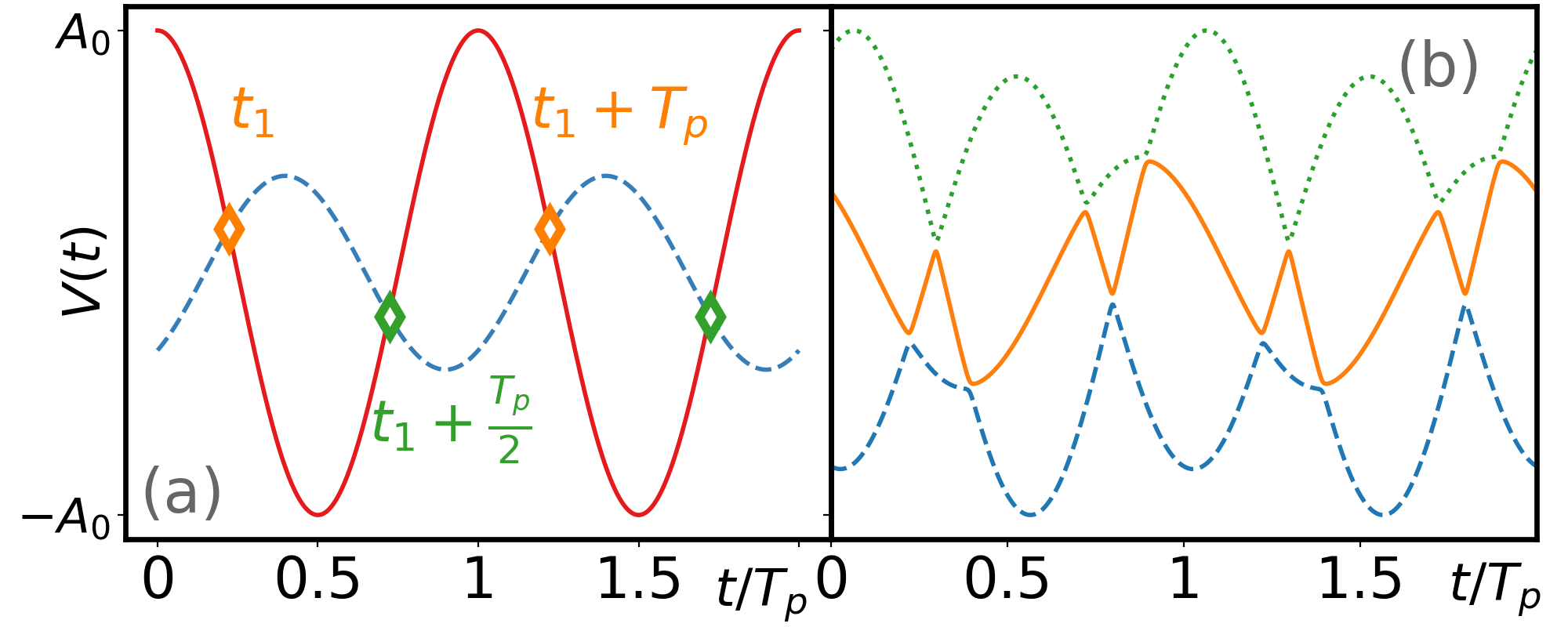}
    \caption{
      (a) Potential energy versus time in a system with two sites.
      The symbols mark the avoided crossings, which are separated by a half phonon period.
      (b) Time-dependent potentials of the three-site semiclassical model with periodic boundary conditions.
      After considering the avoided crossings, there are three possible trajectories if the particle starts from site-0 (dotted), site-1 (dashed), or site-2 (solid).
    }
    \label{fig:FIG1}
  \end{center}
\end{figure}

\textit{One dimension --}
Consider first the Holstein model on two sites.
Although our goal is eventually to treat fully quantum phonons,
consider first highly excited classical phonon states on each site, where the phonon has a large initial
position and momentum ($x$,$p$) drawn from a thermal distribution, or else a quantum coherent state with the same
($x$,$p$).  If the hopping $\bar{t}$ is zero, the problem is exactly solvable.
If the electron is on site $0$, the electronic energy varies as $V_0 (t) \!=\! A_0 \cos ( \Omega t \!+\! \phi _ 0 )$ from the electron-phonon coupling.
If the electron is on site $1$, the behavior is similar, with a different amplitude and phase as shown in Fig.~\ref{fig:FIG1}(a).
If nonzero electron hopping $\bar{t}$ is added, one can treat the problem of a quantum electron in a time-dependent potential due to the phonons.
In the adiabatic limit (small $ \Omega $), with $\bar{t}$ smaller than a typical $V_j$, the crossing of $V_0$ and $V_1$ becomes an avoided crossing.
The electron that initially had most of its amplitude on site $0$ hops to site $1$ during the avoided crossing.
If one assumes the functional form of $V_0$ and $V_1$ continues for all time,
the electron hops from site to site every time interval $T_p/2$ [as shown in Fig.~\ref{fig:FIG1}(a)], where $T_p$ is the phonon period, $\Omega T_p \!=\! 2 \pi $.
Note that this is already a violation of the Boltzmann diffusion hypothesis,
because the system does not lose memory of its initial state.
Even in the long time limit, the electron probability is not spread equally on both sites,
but rather retains memory of its initial state, alternating sites like clockwork, forever.
Even in the fully quantum 2-site problem, solved numerically exactly with no approximations~\cite{si},
similar behavior is obtained for certain initial conditions.
For example in the superposition of an even-parity and an odd-parity many-body eigenstate,
the expectation of the excess density on site 1 oscillates like a cosine, forever.

If the other parameters are fixed, the typical on-site energies $V_j$ eventually become larger than the hopping $\bar{t}$ as the temperature increases.
Consider a one-dimensional array of sites with periodic boundary conditions at high temperature, with the electron initially on site $0$ with on-site energy $V_0 (t)$.
If the energy $V_1 (t)$ of the site on the right becomes equal to $V_0 (t)$ before the site on the left does,
the first avoided crossing is between sites $0$ and $1$, and the electron will hop to the right.
Perhaps unexpectedly, if the electron’s first hop is to the right, all subsequent hops are also to the right.
The reason is that if the electron hops right at time $t_1$, the next avoided crossing that will allow it to
hop right again occurs at a random time (depending on phonon amplitudes and phases)
somewhere in the interval $[t_1 , t_1 \!+\! T_p / 2 ]$. 
But the next avoided crossing that would allow it to hop left 
is exactly at time $ t = t_1 \!+\! T_p / 2$, at the very end of the interval.
The hop at the end of the interval occurs later than the one in the interior.
Similarly, if the first hop is to the left, which happens half of the time, all subsequent hops will be to the left.
This is illustrated in Fig.~\ref{fig:FIG1}(b) in a periodic chain with 3 sites.
From all three different initial states, the particle moves in constant direction.
This behavior is completely different from diffusion.
Instead of $\langle x ^ 2 \rangle \! \sim \! t $, it increases as $t^2$.
The average electron velocity is tied to the phonon frequency:
it is a factor of order unity (= 4) times the lattice spacing divided by the phonon period.
We call this approximation in which the electron hops to the neighboring site with the first avoided crossing the semiclassical approximation.


\begin{figure}[b]
  \begin{center}
    \includegraphics[width=0.48\columnwidth]{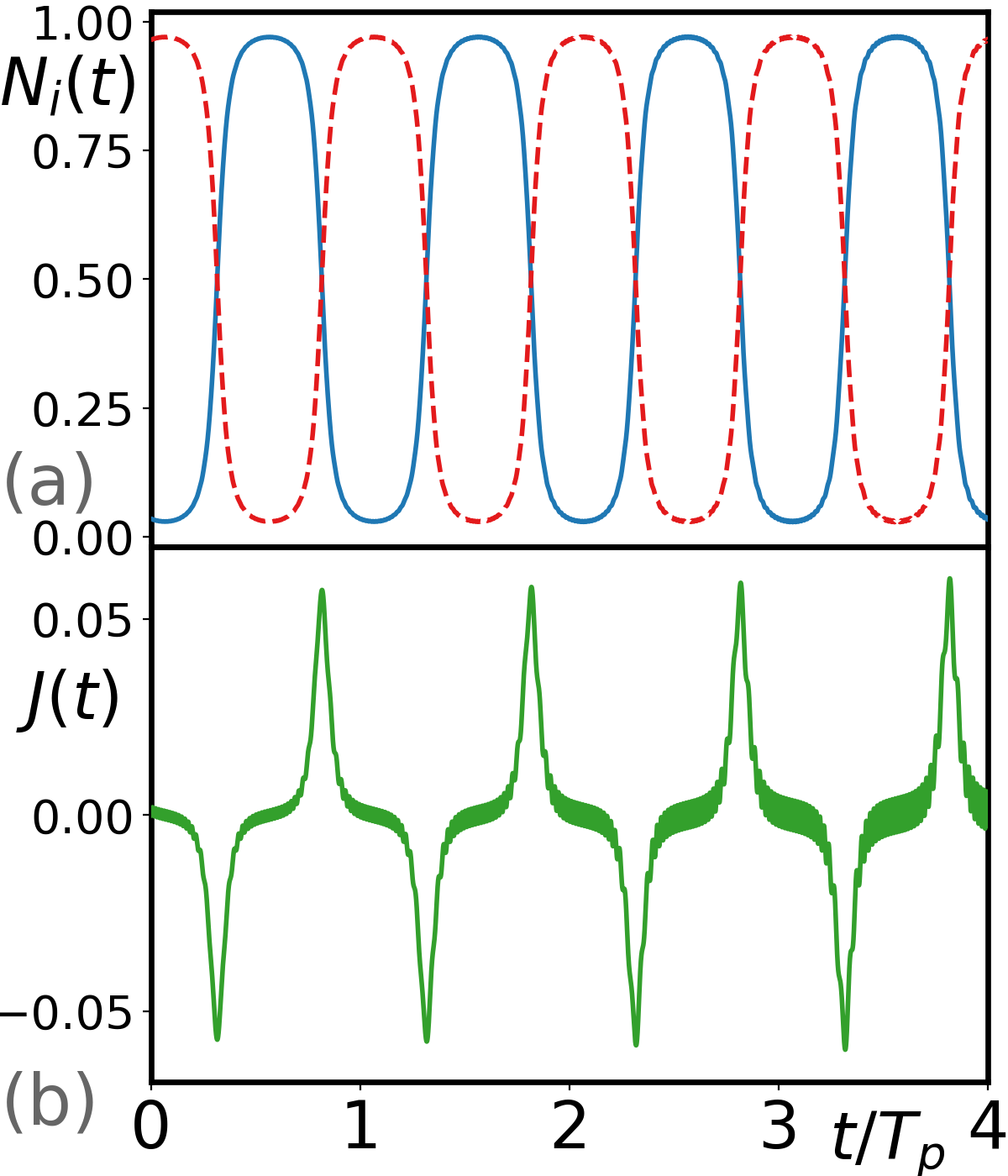}
    \includegraphics[width=0.48\columnwidth]{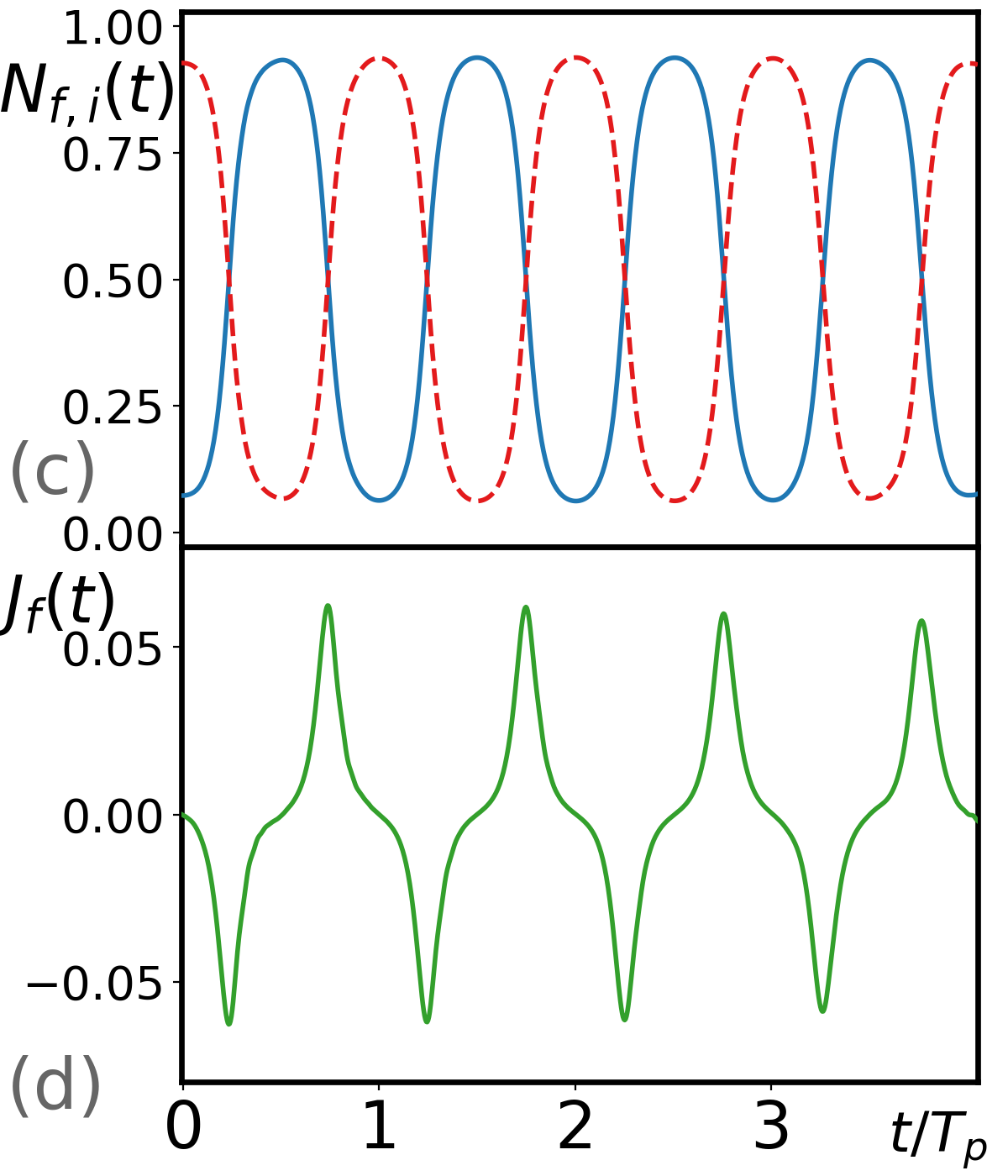}
    \caption{
      (a) Particle and (b) current densities versus time in a two-site system solved by QE approach. 
      (c) Particle and (d) current densities versus time in a two-site system solved by the fully quantum (FQ) approach with phonon coherent state $\vert\alpha\vert\!=\!7.5$ and $\varphi_{0(1)}\!=\!0(\pi)$.
      The system parameters are set to $g\!=\!0.4\bar{t}$ and phonon frequency $\Omega\!=\!0.3\bar{t}$ and $N^\text{max}_p\!=\!120$ is used in the simulation.
    }
    \label{fig:FIG2}
  \end{center}
\end{figure}

If a more accurate calculation, including fully quantum phonons and electron, confirms that the semiclassical approximation
has some validity at high temperatures, this would require a change in
the Boltzmann diffusion scenario.
We will see below that more accurate calculations do provide some confirmation.
First, as an intermediate step, we treat the itinerant particle as a quantum wavefunction and keep the phonons as local time-dependent potentials whose average amplitude increases with temperature $T$ as $A\!\sim\!\frac{g}{\Omega}\sqrt{\frac{T}{m_e}}$ where $m_e$ is the mass of electron.
In short, the last two terms in Eq.~\eqref{eq:holstein} are replaced by $\sum_iV_i(t)n_i$.
We call this treatment the quantum electron (QE) approach.
In this approximation, the initial wavefunction is (Floquet) periodically driven with period $T_p$.
The dynamics of the itinerant particle is simulated~\cite{detail1} by the time-dependent Sch\"{o}dinger equation, $id\Psi/dt\!=\!H(t)\Psi$.
Figures~\ref{fig:FIG2}(a) and~\ref{fig:FIG2}(b) show the time evolution of the particle and current densities in a two-site system.
On each site, the particle density oscillates over time with period $T_p$, with small components at higher frequencies.
The current density on the link shows a narrow peak as the particle density switches from site to site during the avoided crossing.
The results agree with the semiclassical approach, which states that the particle jumps between the two sites repeatedly and stays at each site for half a phonon period.

Finally, we consider the Holstein model of Eq.~\eqref{eq:holstein}
without any approximation, so that both the phonons and the electron are treated quantum mechanically.
The eigenstates and dynamics can be simulated accurately by the time-dependent Lanczos method~\cite{avella2013,zhang2010,lai2016,bonca1999,kogoj2016,hochbruck2006,moler2003}.
Here, the temperature and the coupling are two independent variables.
Due to the unbounded Hilbert space of local phonons, the system size is restricted to be small and a cutoff on total phonon quanta $N^\text{max}_p$ is introduced.
We call the numerically exact complete quantum mechanical treatment the fully quantum (FQ) approach.
We construct the initial state of the quantum phonons by using
phonon coherent states~\cite{shankarr.1994},
\begin{equation}
  \vert\alpha\rangle=\sum_{n}\frac{ \vert\alpha\vert^n e^{in\varphi}}{\sqrt{n!}}\vert n\rangle\;,
\end{equation}
where $\vert\alpha\vert^2\!=\!\langle a^\dagger a\rangle\!\approx\!k_BT/\hbar\Omega$ represents the temperature, 
and the wavefunction of the itinerant particle is taken to be an eigenstate (usually the ground state) of the Hamiltonian with effective local potentials determined from the initial phonon wavefunction~\cite{si}.
Figures~\ref{fig:FIG2}(c) and~\ref{fig:FIG2}(d) present the time evolution of particle and current densities from this given initial state in a two-site system.
The maximum particle density oscillates from site to site and the current density shows a narrow peak as the particle rapidly switches from site to site.
In summary, for a two-site system, the itinerant particle stays at each site for half a phonon period before jumping to the other site in all three different approaches.  This microscopic dynamics is not diffusive.

\begin{figure}[t]
  \begin{center}
    \includegraphics[width=\columnwidth]{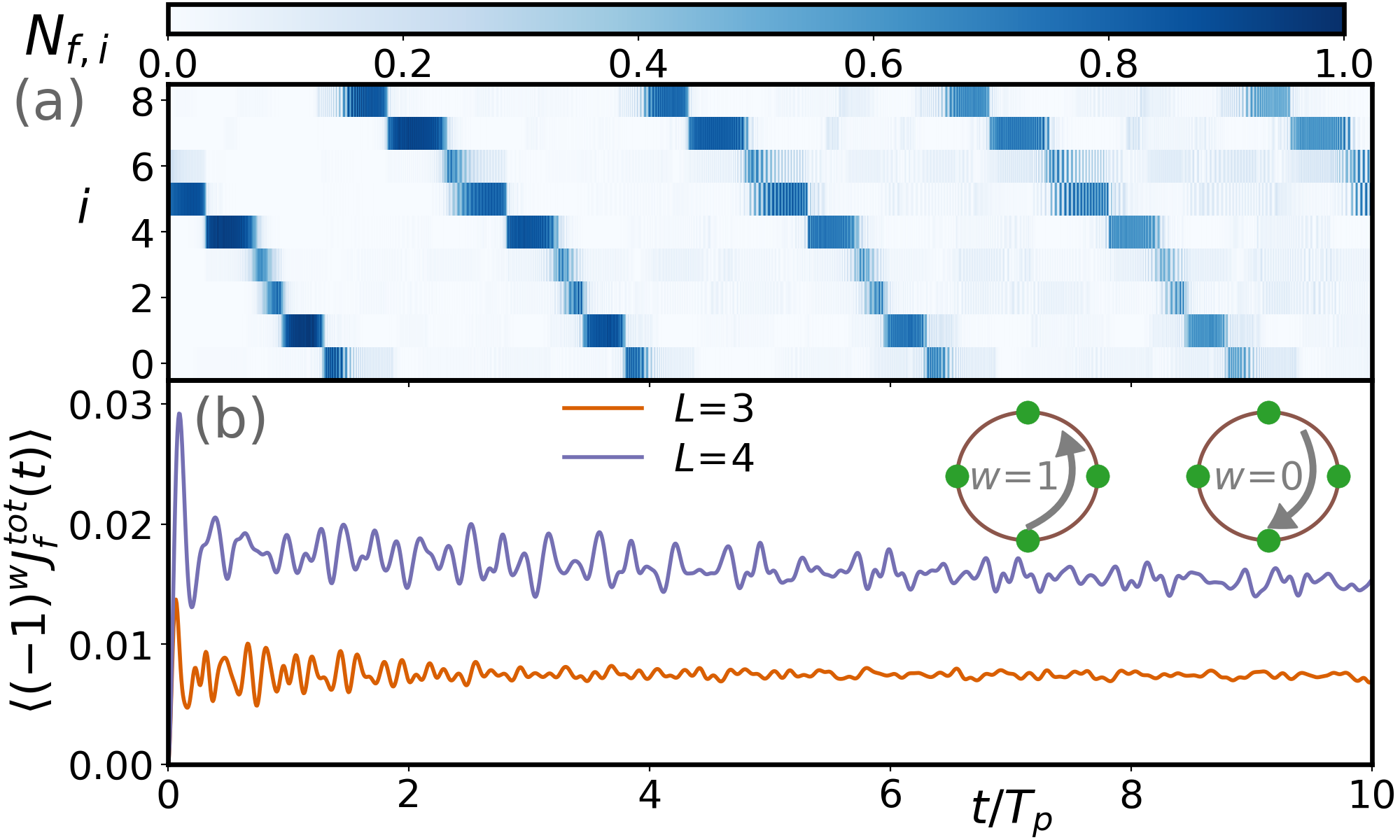}
    \caption{
      (a) Particle density versus time from one realization of the QE model in a 9-site system where the phonon period is set to $T_p\!=\!100t_0$.  The potential amplitudes (phases)
are randomly generated from a thermal (uniform) distribution.  The itinerant particle is initially in the ground state of the Hamiltonian potential.
      (b) The ensemble average of total current on all bonds versus time from the numerically exact FQ simulation with $10^3$ realizations for $L\!=\!3$, and $4$ sites.
      If the particle travels counterclockwise, the current density is multiplied by $-1$ as shown in the inset ($L\!=\!4$).
      The phonon frequency and electron-phonon coupling are set to $\Omega\!=\!0.4\bar{t}$ and $g\!=\!0.3\bar{t}$ respectively.
      The amplitude (phase)
of the coherent state $\vert\alpha \rangle$
is generated randomly from the thermal (uniform) distribution,
      and the maximum number of phonon quanta $N_p^\text{max}\!=\!20L$ is used.
    }
    \label{fig:QE-Coherent-Ensemble}
  \end{center}
\end{figure}

As discussed above, in a one dimensional periodic chain,
the itinerant particle travels in a constant direction given by the initial conditions in the semiclassical approximation.
Figure~\ref{fig:QE-Coherent-Ensemble}(a) shows one realization from the QE model that follows this prediction for a long time.
In a FQ simulation shown in Fig.~\ref{fig:QE-Coherent-Ensemble}(b) starting from phonon coherent states,
the total current retains its initial sign and does not decay to zero
after ten phonon periods (about 36 hops), in the thermal ensemble average.
This indicates that there is a non-zero circulating current in the periodic chain in either one of the directions: clockwise or counterclockwise.
In a diffusion picture, this quantity would decay rapidly to zero
as the particle has an equal chance to move right or left.

The simple semiclassical model assumes that the electron jumps at the first avoided crossing.
This is an oversimplification, in that in a random ensemble, sometimes the
second avoided crossing is only very slightly later, and it would be expected that part
of the quantum amplitude moves in each direction.
In contrast, in both the QE and FQ simulations,
there is no assumption that the particle jumps at the first avoided crossing,
nor that the particle starts from a single site.
All phonon initial conditions are included in the ensemble average as the amplitudes are drawn from a thermal distribution and the phase chosen at random.
Even in the more accurate QE and the exact FQ simulations, the persistence of the current
and lack of diffusion first argued from the simple semiclassical model persist.

The particle is expected to follow the semiclassical prediction
only in certain parameter regimes.
First, Landau-Zener tunneling should be weak (the trajectory should usually not change color in Fig. 1b), which gives ~\cite{landau1932,zener1932,si}
\begin{equation}
  \frac{g\Omega\langle a^\dagger a\rangle^{1/2}}{\bar{t}^2}\ll1\;.
\end{equation}
An implicit assumption of the semiclassical and QE models is that the amplitude and phase of the phonon oscillation is not affected when the electron hops on and off a site.  This approximation is accurate if
\begin{equation}
  g\ll\Omega\langle a^\dagger a \rangle^{1/2}=\Omega\vert\alpha\vert\;.
\end{equation}
This second condition is satisfied at higher temperatures that excite more phonon quanta, and at weaker elextron-phonon coupling $g$.
Both conditions are tested numerically and presented in the Supplemental Material~\cite{si}.

\begin{figure}[b]
  \begin{center}
    \includegraphics[width=\columnwidth]{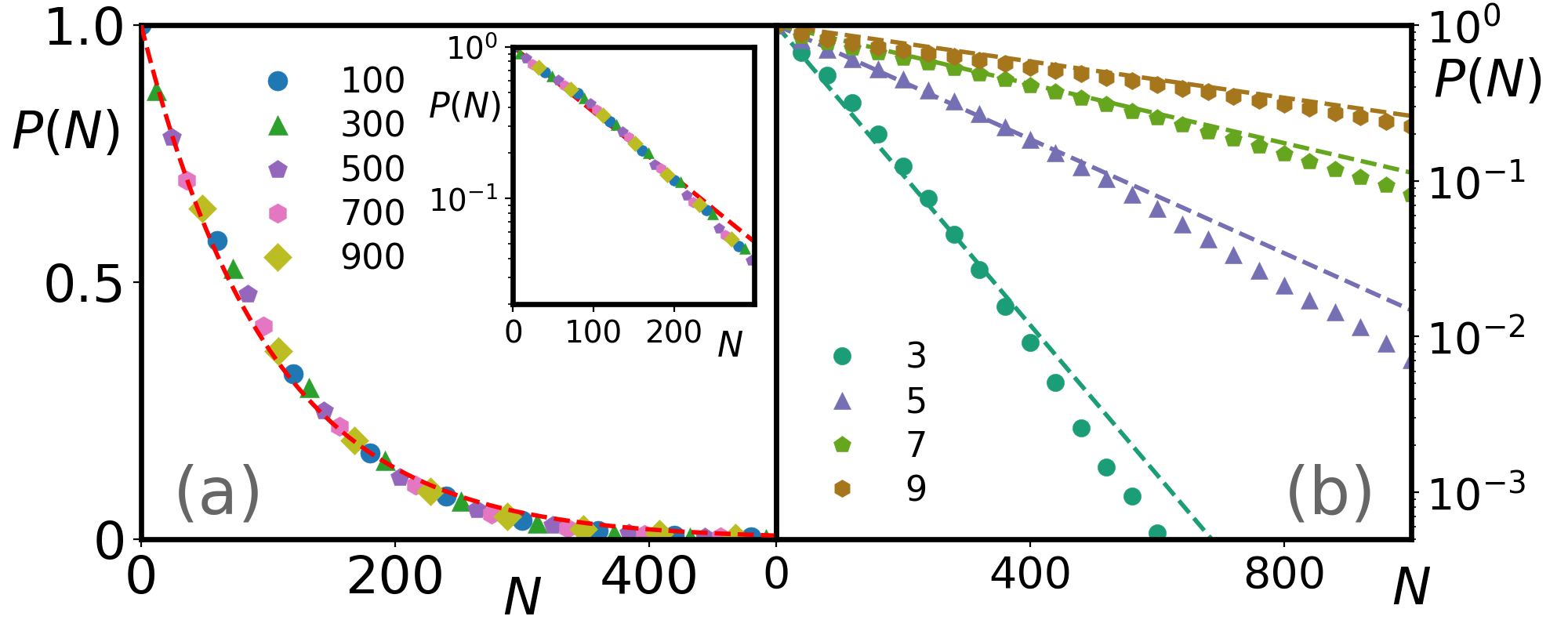}
    \caption{
      (a) Integral of probability, $P(N)$, in a 3-leg ladder system of different lengths $L_x$.
      The inset shows an exponential fit.
      (b) The same plot with different number of rungs $L_y$. The system length is $L_x\!=\!100$.
    }
    \label{fig:ladder}
  \end{center}
\end{figure}


\textit{Ladders --}
In the one dimensional chain with periodic boundary conditions, the particle travels through the entire chain and back to its starting point in the semiclassical model.
For a ladder system ($L_x\!\gg\!L_y$), the particle
need not visit all lattice sites, and a repeating closed trajectory on a square lattice
can contain as few as 4 sites.
In the semiclassical approximation, in any finite system, the particle always traces a closed trajectory
that is repeated periodically in the future and past.
Moreover, the particle will not trace the same bond in the same direction more than once
before closing the trajectory.
Thus, the maximum number of steps to close a trajectory is $N^\text{max}_s\!=\!2N_\text{bonds}$ where the factor $2$ is counting both directions and $N_\text{bonds}$ is total number of bonds in the finite system.
The above two statements are numerically verified in the semiclassical model, and
simulations from the QE model draw
similar conclusions and are presented in the Supplemental Material.
In order to give a more quantitative analysis of the dynamics,
the number of jumps in the closed trajectory, $N_s$, is simulated for more than $10^5$ realizations
of the initial conditions for the semiclassical model.
We aim to answer whether the particle will have a finite trajectory, with probability 1, in the thermodynamic limit.
Here, the probability of taking more than $N$ jumps to close a trajectory is used to quantify the behavior of the itinerant particle.
Explicitly, we define
\begin{equation}
  P(N)\!=\!\int^{N^\text{max}_s}_N p(N_s)dN_s\;,
\end{equation}
where $p(N_s)$ is the probability of taking $N_s$ jumps to close the trajectory.
First, by keeping number of rungs $L_y$ fixed and expanding the length of the chain $L_x$, the results in Fig.~\ref{fig:ladder}(a) show that $P(N)$ decays exponentially and follows the same scaling regarding different lengths.
The results of increasing the number of rungs $L_y$ in Fig.~\ref{fig:ladder}(b) as the system approaches a two-dimensional (2D) square lattice show that the exponent decreases as $L_y$ increases.

\begin{figure}[t]
  \begin{center}
    \includegraphics[width=\columnwidth]{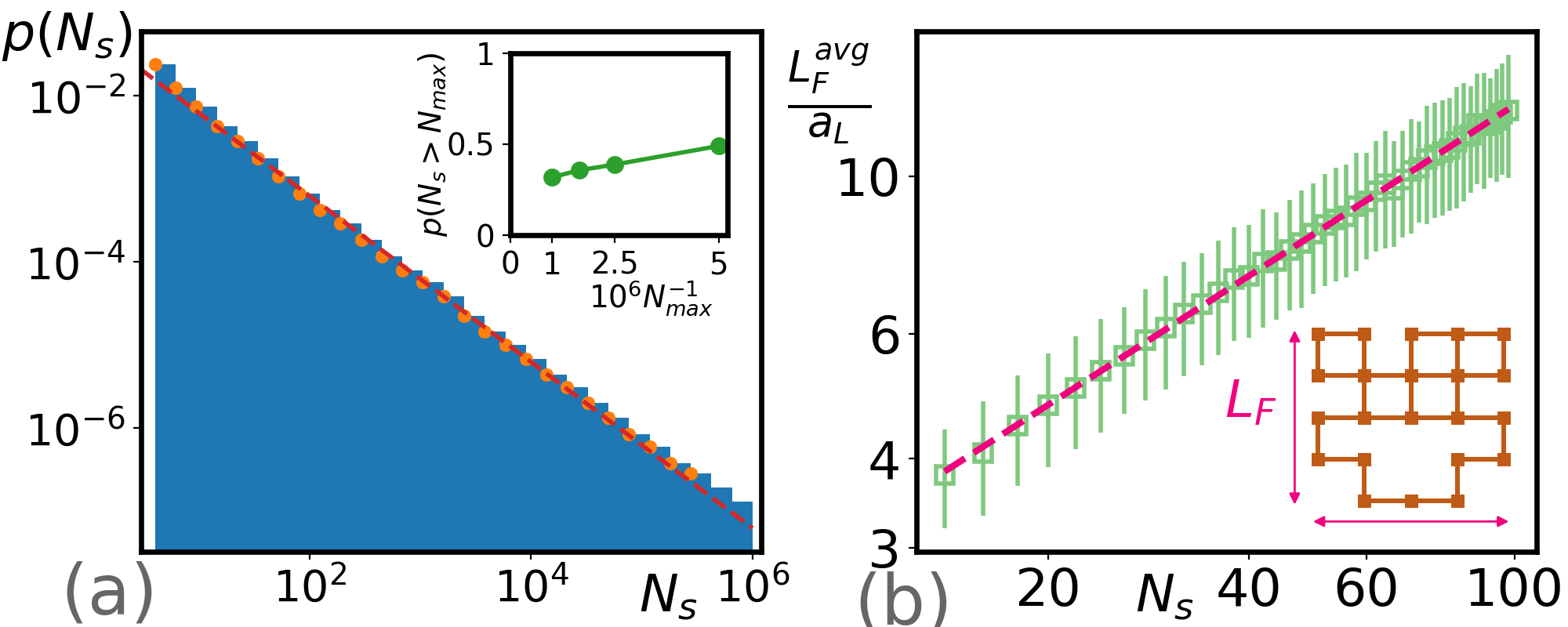}
    \caption{
      (a) The probability $p(N_s)$ on an infinite square lattice with upper limit $N_s\leq10^6$ and more than $10^5$ realizations.
  The inset shows the probability that the trajectory has not closed after $N_\text{max}$ steps.
      (b) The average box size of the closed trajectories versus $N_s$. The error bars are standard deviations.
      The inset shows one of many realizations with $N_s\!=\!24$ and $L_F\!=\!4a_L$ where $a_L$ is lattice constant.
    }
    \label{fig:Square}
  \end{center}
\end{figure}

\textit{Square lattices --}
Next, we consider the dynamics on an infinite two dimensional square lattice
with a cut-off maximum ($N_\text{max}$) of $10^6$ jumps in the semiclassical model.
Unlike the ladder systems, the probability $p(N_s)$ decays as power law as shown in Fig.~\ref{fig:Square}(a), rather than exponentially.
%
%
This power law decay indicates that the system is critical in 2D.
The inset show the probability of having more than $N_\text{max}$ jumps as a function of $N_\text{max}$.
The finite size scaling extrapolation of $N_\text{max} \rightarrow \infty$ appears to show that a finite fraction of the trajectories in 2D, approximately $0.28$, are infinite and never close.
The critical behavior suggests that the trajectory is fractal.
Let $L_F$ be the edge length of the smallest square box that can enclose a given trajectory.
Figure~\ref{fig:Square}(b) shows that the relation between the average $L_F$ and the number of jumps between $N_s\!=\!14$ and $98$.
The average size of the square scales as $L_F\!\sim\!N_s^{\phi}$ with non-integer exponent $\phi\!=\!0.6\!\pm\!0.003$, which indicates that the trajectory is fractal.
The fractal geometry found here is qualitatively similar to the one in the integer quantum Hall effect~\cite{wang2015,dean2013,trugman1989} where the particle travels in a random potential in a magnetic field.

\textit{Conclusions --}
The microscopic dynamics of an electron in the Holstein model of electron-phonon coupling at high temperatures is studied in two approximations, and in a numerically exact fully quantum approach.
The results of these approaches agree with each other qualitatively, at least for intermediate times.
Surprisingly, the dynamics are not diffusive.
In a periodic one-dimensional chain, the electron will move in a single direction, return to its starting point, and loop repeatedly.
On ladder systems, the particle travels in a closed loop, with long trajectories exponentially rare.
On a 2D lattice, the system is critical.  The trajectories are fractal, long trajectories have a power law probability, and a finite fraction of the trajectories are infinite (never close).

\acknowledgments{
The authors thank Eli Ben-Naim, Janez Bon\u{c}a, Peter Prelov\u{s}ek, and Daniel Trugman for fruitful discussions.
This work was performed, in part, at the Center for Integrated Nanotechnologies,
an Office of Science User Facility operated for the U.S. Department of Energy (DOE) Office of Science.
Los Alamos National Laboratory (LANL), an affirmative action equal opportunity employer,
is managed by Triad National Security, LLC for the U.S. Department of Energy’s NNSA, under contract 89233218CNA000001.
This work was also supported by LANL LDRD.  Computational resources were provided by the LANL Institutional Computing Program.
}


\begin{thebibliography}{18}
\bibitem[{det()}]{detail1}
  \bibinfo{note}{The initial state is choosen as the ground state of the
    Hamiltonian at time $t\!=\!0$ and the real time dynamics is solved by
    fourth-order Runge-Kutta with time step $10^{-3}t_0$ where the time unit is
    $t_0\!=\!\hbar/\bar{t}$.}
\end{thebibliography}

\end{document}


\title{Supplemental Material: Absence of Diffusion and Fractal Geometry in the Holstein Model at High Temperature}

\author{Chen-Yen Lai}
\author{S. A. Trugman}
\affiliation{Theoretical Division, Los Alamos National Laboratory, Los Alamos, New Mexico 87545, USA}
\affiliation{Center for Integrated Nanotechnologies, Los Alamos National Laboratory, Los Alamos, New Mexico 87545, USA}

\date{\today}

\maketitle


\begin{figure}[b]
  \begin{center}
    \includegraphics[width=\columnwidth]{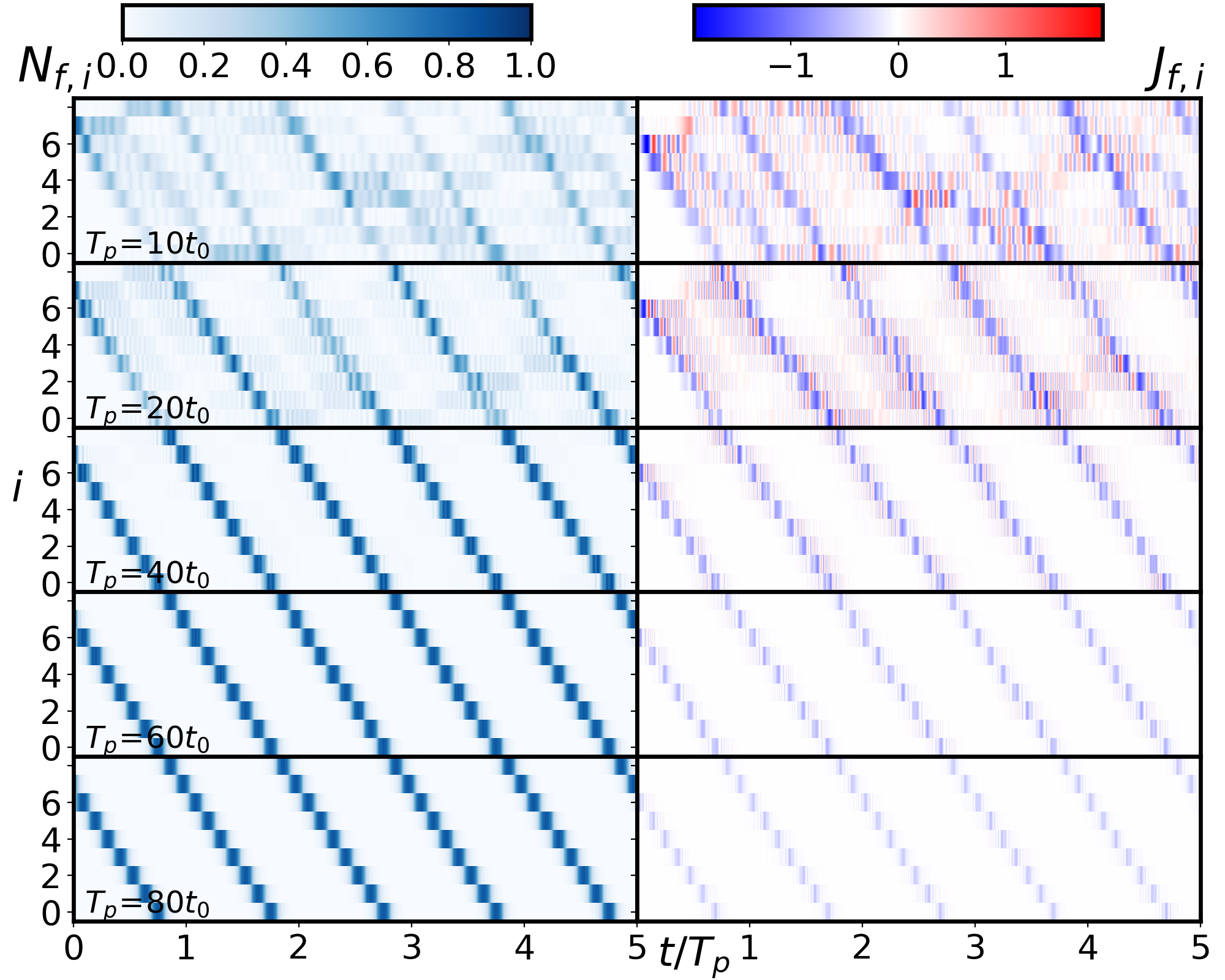}
    \caption{
      The particle density (left) and current (right) versus time for the quantum electron model in a 9-site system with periodic boundary conditions.
      The phonon period is labeled in the figure, the amplitude of potential is set to $A_i\!=\!12\bar{t}$ on all sites and the phase of the potential is taken to be $\phi_i\!=\!2i\pi/9$ for $i\!=\!0,\cdots,8$.
      The particle has maximum density on site 7 initially and the maximum moves from site 7 to site 6, traveling all the way to site 0 and crossing the periodic boundary.
      The initial state is the ground state of the corresponding Hamiltonian.
    }
    \label{fig:QE-Omega}
  \end{center}
\end{figure}

\begin{figure}[b]
  \begin{center}
    \includegraphics[width=\columnwidth]{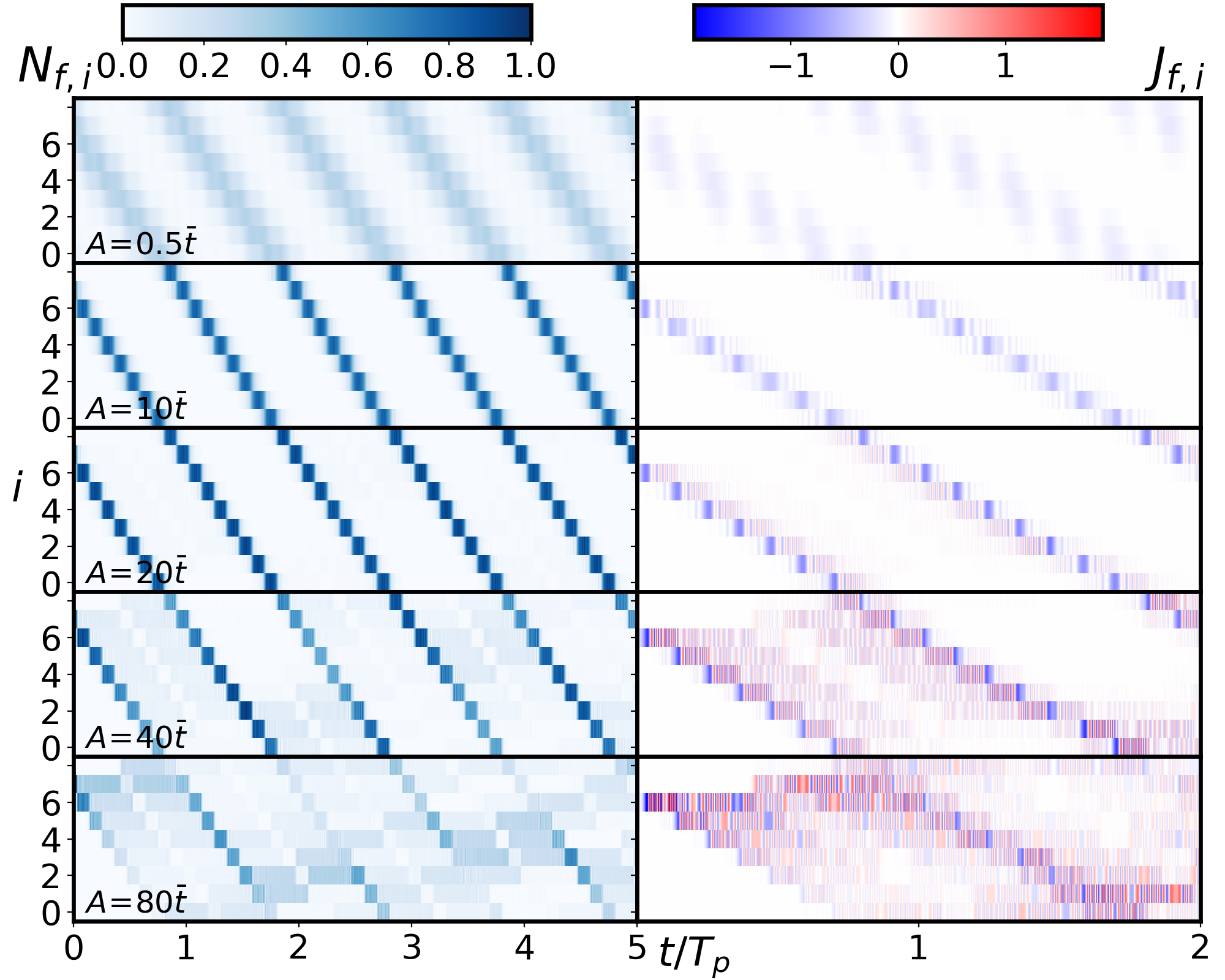}
    \caption{
      The particle density (left) and current (right) versus time for the quantum electron model.
The conditions are the same as Fig. S1, except that here the phonon amplitude is varied, with the phonon period fixed at $T_p\!=\!60t_0$.
    }
    \label{fig:QE-Amplitudes}
  \end{center}
\end{figure}

\section{Quantum Electron}

In this section, we elaborate on the quantum electron (QE) model.
The electron
is described by a quantum wavefunction
$\Psi(j,t)$ and evolves by the time-dependent Sch\"{o}dinger equation,
$i \partial_t\Psi(t) \!=\! H_\text{QE}(t)\Psi(t)$,
where the time dependent Hamiltonian is given by
\begin{equation}
  H_\text{QE}(t)=-\bar{t}\sum_{\langle ij \rangle}(c^\dagger_ic_j+h.c.) + \sum_i V_i(t)c^\dagger_ic_i\;.
\end{equation}
Here, the time dependent potential is $V_i(t)\!=\!A_i\cos(\frac{2\pi}{T_p}t\!+\!\phi_i)$ with phonon period $\Omega T_p\!=\!2\pi$.
The amplitude and phase on each site can be random.
The particle density can be calculated from
$N_{j}(t)\!=\!\langle c^\dagger_jc_j\rangle$ as well as the current density $J_{j}(t)\!=\!2\text{Im}\langle c^\dagger_{j}c_{j+1} \rangle$.
Although the phonon is still described as a simple harmonic potential, the itinerant electron is quantum and does not stay on a single site.
In the semiclassical model, we ignored the possibility that two avoided crossings may occasionally be so close to each other in time that only part of the electron amplitude moves towards the first one.  This splitting is, however, included in the QE model.
First, we check whether the motion of the electron follows the semiclassical picture in certain parameter regimes.
The assumption in the semicalssical model is that the electron moves adiabatically as two potentials cross each other.
In the semiclassical model, the electron-phonon coupling and the effective temperature are not independent variables and together determine the amplitude of the potentials as $A\!\sim\!g\sqrt{T/\Omega},$ where $T$ is the temperature.
In the high temperature regime, the amplitudes grow large and eventually $A\!\gg\!\bar{t}$.
The condition for adiabatic evolution is $\Omega A\!\ll\!\bar{t}^2$, which will be discussed later in detail (see also Eq. (3) in the main text), so the phonon period $T_p$ should be large (small $\Omega$) for the semiclassical model to be accurate.

\subsection{Periodic Chain}
First, in a two-site system shown in
Fig.~1(a) of the main text, the results show that the maximum particle density shifts from site to site as expected in the semiclassical model.
The current density on the bond, shown in
Fig.~2 of the main text,
shows a narrow peak near the time when the avoided crossing occurs.
Next, in a large system, we design both the amplitudes and phases of the potentials such that the particle is expected to spend the same time on each site and travel through the entire system in a single phonon period.
In Fig.~\ref{fig:QE-Omega} (top panel), the simulation shows that the dynamics of the more accurate QE model for a small phonon period (large $\Omega$) do not accurately follow the prediction of the simple semiclassical picture.
As compared to the lower panels, the particle in the upper $T_p\!=\!10t_0$ panel exhibits only fainter "ghosts" of the semiclassical trajectory.
As the phonon period increases, the maximum particle density starts to follow the predicted trajectory and the current density shows a sudden peak as the maximum particle density shifts from site to site in the $T_p\!=\!40t_0$ case.
In the slow phonon regime, $T_p\!=\!80t_0$, the particle density seems to follow the predicted trajectory well.
Figure~\ref{fig:QE-Amplitudes} shows a simulation for different amplitudes $A$.
When $A$ is small, the particle density is rather uniform and the current density does not have narrow peak as expected, but there is a shadow of the behavior that is seen below.
As the phonon amplitude $A$ increases, both the particle density and the current density behave like the one in semiclassical model.
For the largest $A$, the trajectory becomes less distinct.
This is because the adiabatic dynamics breaks down in the large $A$ regime, which suggests that the temperature (electron-phonon coupling) can not be too high (strong).
We will discuss and elaborate on this effect in a later section.

\begin{figure}[t]
  \begin{center}
    \includegraphics[width=\columnwidth]{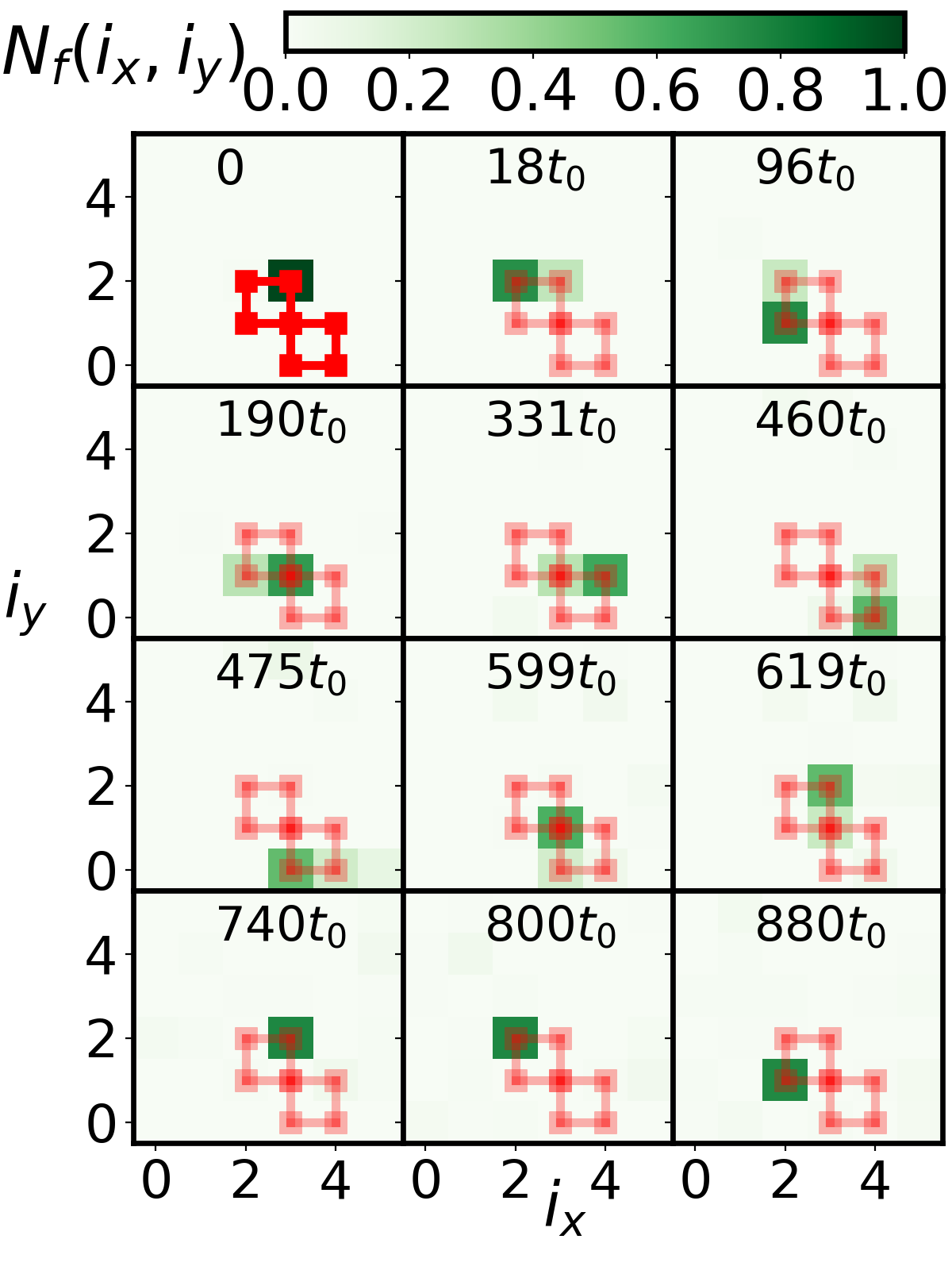}
    \caption{
      One realization from the QE model on a square lattice ($L_x\!=\!L_y\!=\!6$) shows the particle density at different times.
      Around $720t_0$, the particle arrives at its starting point and continues to make the same jumps in the future.
      The potentials have constant phonon period $T_p\!=\!80t_0$, random phases and random amplitudes with average $\bar{A}\!=40\bar{t}$.
    }
    \label{fig:QE-Square}
  \end{center}
\end{figure}

\subsection{Square Lattice}
Figure~\ref{fig:QE-Square} shows one of many realizations from the QE model on a square lattice.
Periodic boundary conditions are used in both directions.  The particle starts in the ground state of the Hamiltonian.
In the example, by tracking the maximum particle density, the particle primarily visits only $7$ sites as marked in red squares in $t\!=\!0$ panel.
After that, the particle continues to repeat the same closed loop
several times before diffusing elsewhere.

\section{Fully quantum}
Here, we simulate
the dynamics of Hamiltonian, Eq.~(1) of the main text,
using a time-dependent Lanczos method
~\cite{avella2013,zhang2010,lai2016,bonca1999,kogoj2016,hochbruck2006,moler2003}.
Since the phonon number is unbounded in the model and there are many thermally-excited phonons in the system, a large phonon cutoff, $N_p^\text{max}$, is required and that makes the simulation challenging.
The dynamics are simulated from an initial state in which the phonons are in coherent states.
The fully quantum simulation is numerically exact, and in this sense is far superior to the semiclassical or the intermediate accuracy quantum electron simulations.  The fully quantum simulation is, however, limited to small systems.

\subsection{Phonon Coherent State}
Coherent states form an overcomplete basis for phonon excitations that resemble the classical states of a given position and momentum.
A coherent state is given by
\begin{equation}
  \vert\alpha\rangle= e ^ { \frac{- \vert \alpha \vert ^2} {2} } ~ \sum_{n}\frac{\alpha^n}{\sqrt{n!}}\vert n\rangle\;,
\end{equation}
where $\vert n\rangle$ is the occupation number representation and $\alpha=\vert\alpha\vert e^{i\varphi}$ is a complex number. ~\cite{shankarr.1994}

We choose the initial wavefunction to be the product of an electronic wavefunction and a phonon coherent state on each site.  For a system with three sites,
\begin{eqnarray}
  &&\vert\Psi(t=0)\rangle=\sum_{i_F}\psi_{i_F}\vert i_F;\alpha_0, \alpha_1, \alpha_2\rangle\\
  &=&\mathcal{N}^{-1}\sum_{i_F}\sum_{n_0, n_1, n_2}\psi_{i_F}\frac{\alpha_0^{n_0}\alpha_1^{n_1}\alpha_2^{n_2}}{\sqrt{n_0!\,n_1!\,n_2!}}\vert i_F;n_0,n_1,n_2\rangle\;,\nonumber
\end{eqnarray}
where $\mathcal{N}$ is a normalization factor and $i_F$ denotes the location of the itinerant particle.
The initial local phonon number $\langle a^\dagger_i a_i\rangle\!=\!\vert\alpha_i\vert^2$ should be correctly reproduced to check whether the phonon cutoff is large enough.
The effective temperature is related to the average phonon number on each site,
$k_BT\!\approx\!\hbar\Omega\langle a^\dagger a\rangle$.

The initial state of the itinerant particle is chosen to be the ground state of the Hamiltonian with effective potentials calculated from the electron phonon coupling and randomly given phonon coherent states.
The effective local potential can be determined as $V_{i}\!=\!-g\langle i_F\!=\!i;\alpha_0,\alpha_1,\alpha_2\vert (a^\dagger_i+a_i) \vert i_F\!=\!i;\alpha_0,\alpha_1,\alpha_2 \rangle$.
For instance, in a three-site system, the initial state of the itinerant particle is the ground state of
\begin{equation}
  H_\text{eff}=\left(\begin{matrix}
      V_0 && -\bar{t} && -\bar{t} \\
      -\bar{t} && V_1 && -\bar{t} \\
      -\bar{t} && -\bar{t} && V_2 \\
    \end{matrix}\right)\;.
\end{equation}
The particle density, shown in
Fig.~2(c) of the main text, agrees with the semiclassical and the QE model qualitatively with the maximum shifting between two sites as the potentials cross each other.
Fig.~2(d) of the main text
shows the current density on the bond.  It has a narrow peak near the potential crossing as seen in the quantum electron model.
This is exactly the motion predicted in both the QE and semiclassical model.

\begin{figure}[t]
  \begin{center}
    \includegraphics[width=\columnwidth]{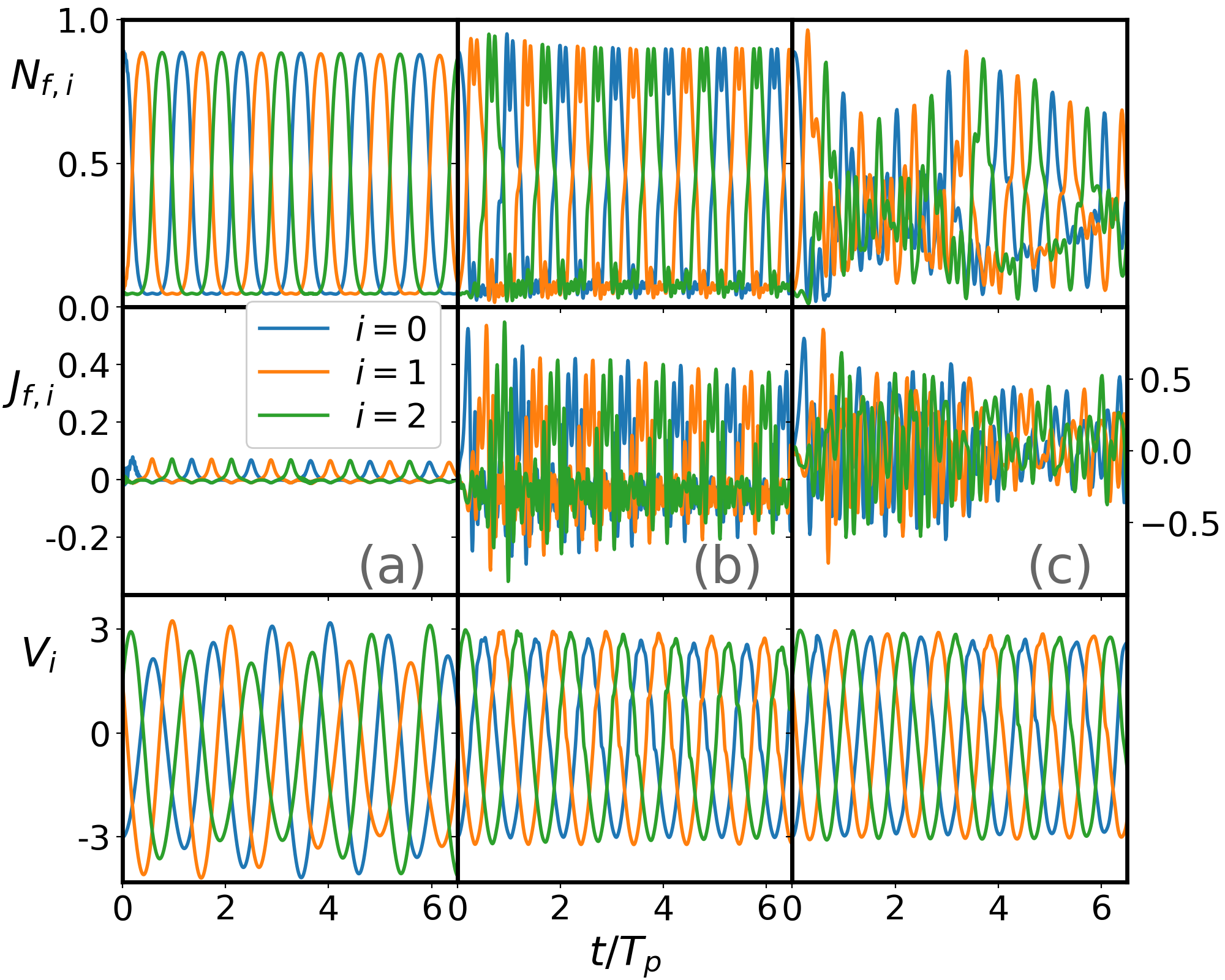}
    \caption{
      The particle density (top), current (middle), and effective potential (bottom) versus time simulated for the fully quantum model for three different phonon frequencies $\Omega/\bar{t}\!=\!0.1$ (a), $0.5$ (b) and $0.9$ (c).
      The coupling is set to $g\!=\!0.3\bar{t}$ and the coherent states have $\forall i\vert\alpha_i\vert\!=\!5$, $\varphi_0\!=\!0$, $\varphi_1\!=\!2\pi/3$, and $\varphi_2\!=\!-2\pi/3$.
      The maximum number of phonon quanta in the simulation is set to $N_p^\text{max}\!=\!120$, here and the following two figures.
    }
    \label{fig:CoherentDesign-OmegaNp}
  \end{center}
\end{figure}

\subsubsection{Landau-Zener tunneling}
An avoided crossing happens when two site potentials cross each other in the presence of tunneling $\bar{t}$ between the sites.
In the adiabatic limit, the particle remains in the instantaneous eigenstate, as assumed in the semiclassical model.
If Landau-Zener tunneling occurs, the particle jumps to the other band, and adiabaticity fails.
The gap at the crossing point is $2\bar{t}$ and the time taken to cross is
\begin{equation}
  \Delta t \!=\!\frac{2\bar{t}}{dV/dt}\!=\!\frac{2\bar{t}}{\Omega V},
\end{equation}
where $\Omega$ is the frequency of the time-varying potential, the phonon frequency.
The Zener formula for the probability of tunneling between bands is
$P\!\approx\!\exp[-\frac{2\bar{t}^2}{\Omega V}]$.
For the semiclassical model to apply, this probability should be small, which gives us the first condition
  $\frac{2\bar{t}^2}{\Omega V}\gg1$.
The amplitude of effective potential in the semiclassical model can be determined from $A\!=\!2g\langle a^\dagger + a\rangle_\text{max}\!=\!2g\langle a^\dagger a\rangle^{1/2}$.
Thus, the criterion for the semiclassical model to apply is
\begin{equation}\label{eq:zener}
  \frac{\bar{t}^2}{\Omega g\langle a^\dagger a\rangle^{1/2}}\gg1.
\end{equation}
The phonon frequency and electron-phonon coupling should be small.
This condition is confirmed for the QE model in Figs.~\ref{fig:QE-Omega} and S2.

For the numerically exact fully quantum simulation shown in Fig.~\ref{fig:CoherentDesign-OmegaNp}, three different phonon frequencies are compared and it is clear that the smallest phonon frequency agrees more closely with the semiclassical model.
For larger phonon frequency, although the effective potential follows the simple harmonic motion, the maximum particle density has strong fluctuations and the current densities on each bond do not show a clear constant direction.
For the smallest phonon frequency
$\Omega\!=\!0.1\bar{t}$, both the maximum particle density and current densities seem to follow better than $\Omega\!=\!0.9\bar{t}$.
For $\Omega\!=\!0.1\bar{t}$ shown in Fig.~\ref{fig:CoherentDesign-OmegaNp}(a), when the effective potentials cross, the maximum particle density shifts from site to site and the current density on each bond shows a narrow peak which agrees qualitatively with the semiclassical and QE model.

\begin{figure}[t]
  \begin{center}
    \includegraphics[width=\columnwidth]{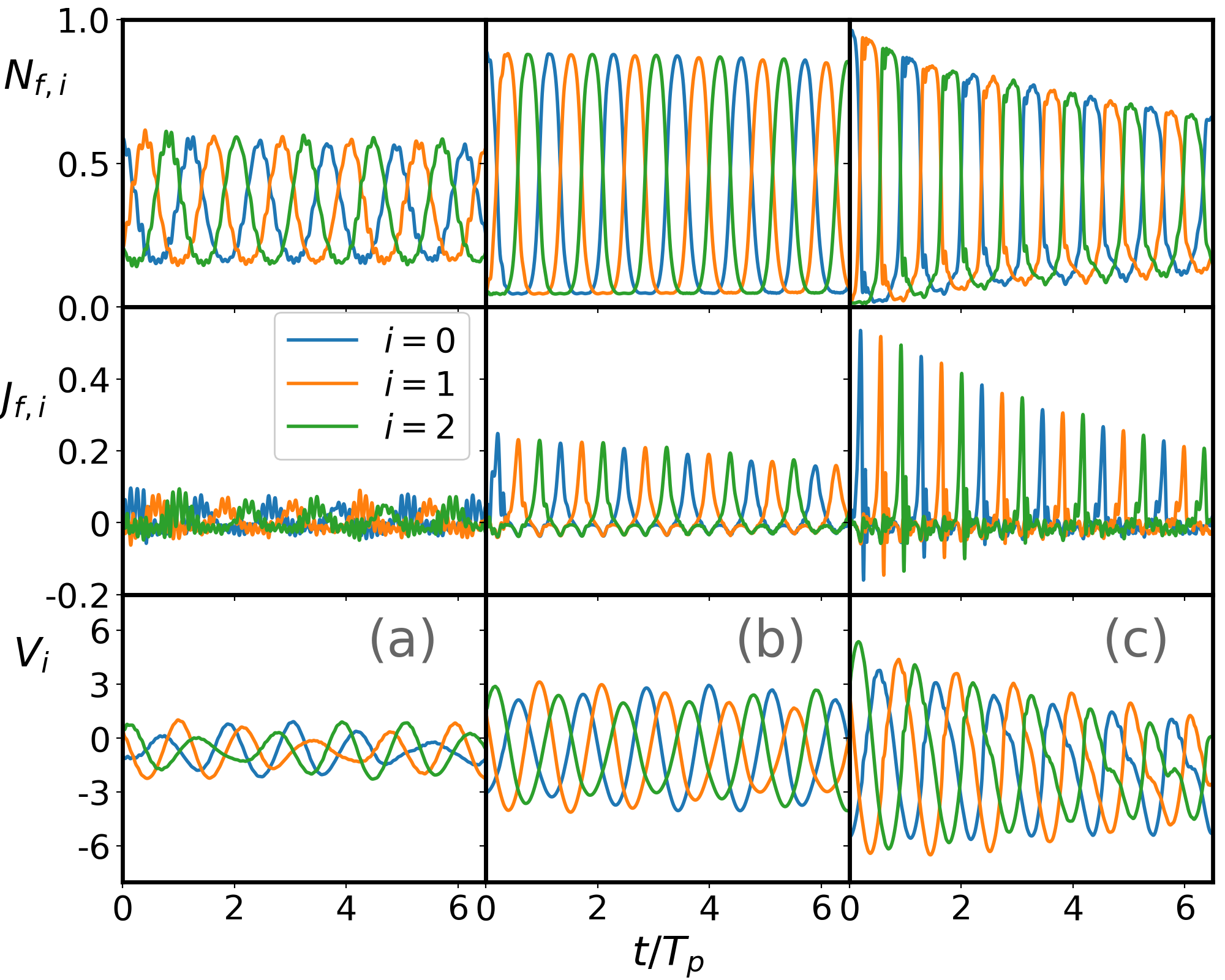}
    \caption{
      The particle density (top), current (middle), and effective potential (bottom) versus time simulated in the fully quantum model for three different amplitudes of the coherent state  $\vert\alpha\vert\!=\!1$ (a), $3$ (b) and $5.5$ (c).
      The frequency is $\Omega\!=\!0.3\bar{t}$, the coupling is set to $g\!=\!0.5\bar{t}$, and the coherent states have phases
$\varphi_0\!=\!0$, $\varphi_1\!=\!2\pi/3$, and $\varphi_2\!=\!-2\pi/3$.
    }
    \label{fig:CoherentDesign-AsNp}
  \end{center}
\end{figure}

The semiclassical model is valid at high enough temperature, where increasing temperature increases the phonon amplitude
$\vert\alpha\vert$, Figure~\ref{fig:CoherentDesign-AsNp}.
For small $\vert\alpha\vert$, although the maximum particle density follows the semiclassical prediction, it is rather uniform compared to larger $\vert\alpha\vert$.
Also, the current densities are broadened in time instead of having a narrow peak.
Increasing $\vert\alpha\vert$ to $3$ is closer to our semiclassical model for the maximum particle density, current densities, and the effective potential.
We have mentioned that the temperature cannot be too high in the QE approach.
This can be seen in Eq.~\eqref{eq:zener}, where the inequality is violated if the number of thermally excited phonons $\langle a^\dagger a\rangle$ grows too large.
If one increases the temperature to a larger value, $\vert\alpha\vert\!=\!5.5$, the results from the fully quantum model support the same conclusion.
The motion is well behaved in the short time regime, but it does not last a long time compared to the $\vert\alpha\vert\!=\!3$ case.

\begin{figure}[t]
  \begin{center}
    \includegraphics[width=\columnwidth]{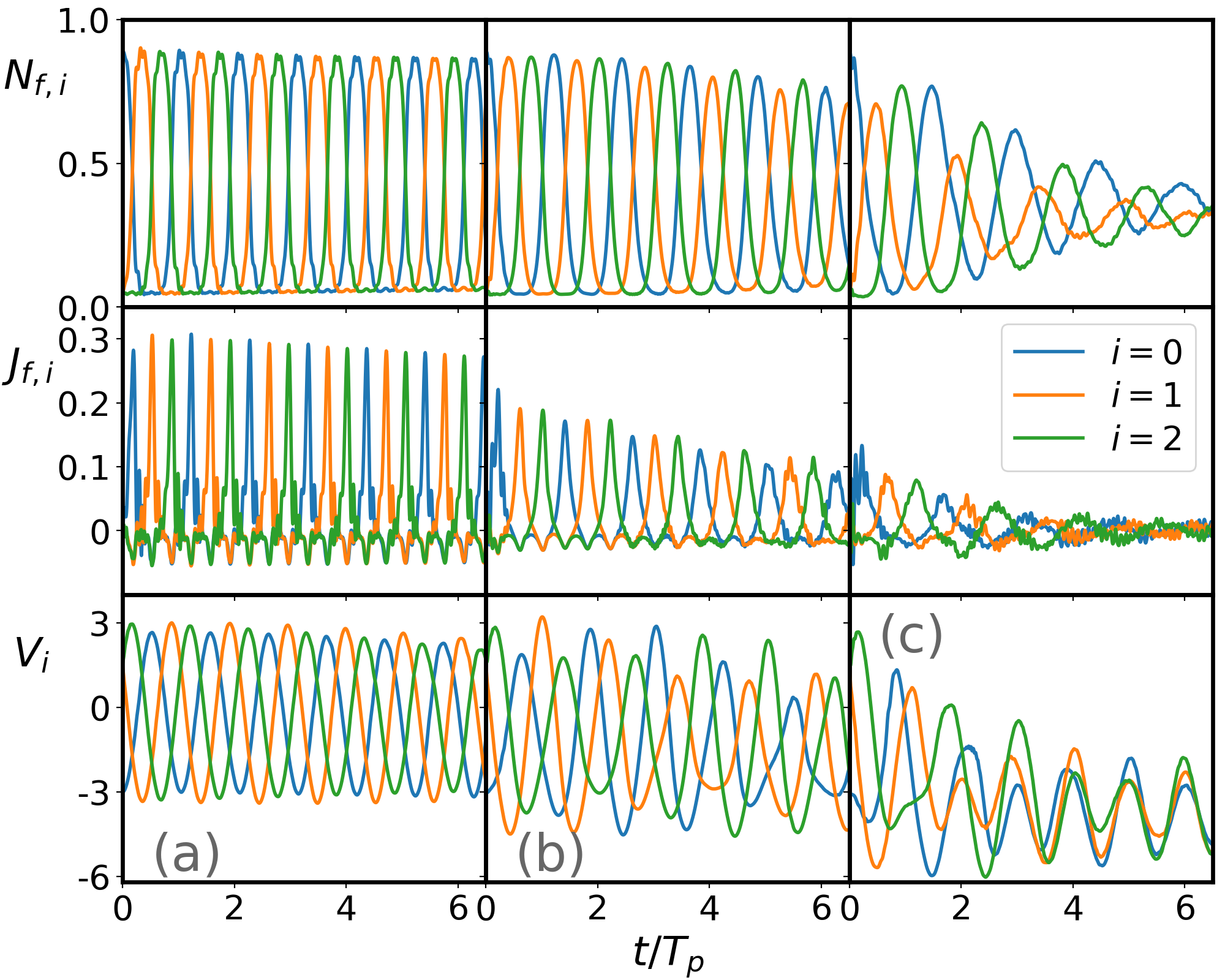}
    \caption{
      The particle density (top), current (middle), and effective potential (bottom) versus time simulated in the fully quantum model for three different combinations of $(g,\vert\alpha\vert)$ in (a) $(0.3, 5.0)$, (b) $(0.6, 2.5)$, and (c) $(0.8, 1.875)$.
In the semiclassical model, the effective potentials would all oscillate between
$\pm3\bar{t}$.
      The coherent state parameters are given by the same $\vert\alpha_i\vert$ $\forall i$, $\varphi_0\!=\!0$, $\varphi_1\!=\!2\pi/3$, and $\varphi_2\!=\!-2\pi/3$.
      The frequency is set to $\Omega\!=\!0.3\bar{t}$.
    }
    \label{fig:CoherentDesign-LambdaNp}
  \end{center}
\end{figure}

\subsubsection{Recoil}
The magnitude of the effective potential in Fig.~\ref{fig:CoherentDesign-OmegaNp}(a) varies with position and time.
This suggests another constraint for the semiclassical model to work: the phonon recoil from the electron-phonon coupling.
A phonon coherent state circles the origin in phase space ($x,p$) with radius $\vert\alpha\vert$.
If an electron jumps onto the site, the center of the orbit shifts by $ \Delta x = g / \Omega$.
For the semiclassical model, which ignores this shift, to apply, the shift should be small compared to the radius of the orbit,
\begin{equation}\label{eq:recoil}
  \frac{g}{\Omega}\ll\langle a^\dagger a \rangle^{1/2}=\vert\alpha\vert\;.
\end{equation}

Fig.~\ref{fig:CoherentDesign-LambdaNp} shows the fully quantum simulation
for different combinations of $g$ and $\vert\alpha\vert$.
The figure on the left follows the semiclassical approximation rather well.
Moving to the right, the chosen parameters violate the recoil inequality Eq. (S7) more and more strongly.
And indeed, the dynamics can be seen to track the semiclassical model less and less well.  (Note that the chosen parameters are not testing the Zener inequality, as they all leave the LHS of Eq. (S6) unchanged.)

\begin{figure}[b]
  \begin{center}
    \includegraphics[width=0.8\columnwidth]{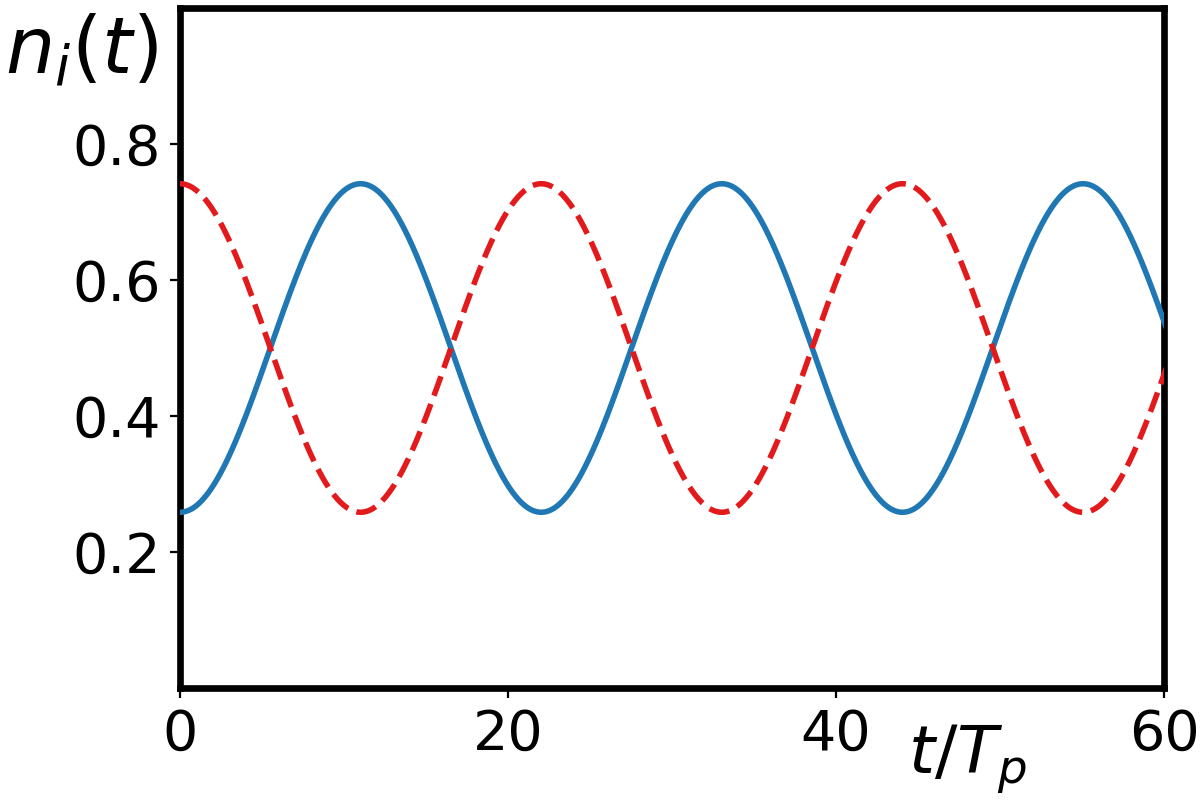}
    \caption{
      The particle density on site-$0$ (solid) and $1$ (dashed) versus time from a pair of excited states in a two-site system.
      The phonon frequency and electron-phonon coupling are set to $\Omega\!=\!g\!=\!0.4\bar{t}$.
    }
    \label{fig:Excited-TwoStates}
  \end{center}
\end{figure}


\subsection{Excited States}
The fully quantum dynamical simulations are numerically demanding.
The size of the Hilbert space can be reduced by working in momentum space, and noting that the
$\qb\!=\!0$ phonon mode can be omitted.  This is because it couples only to the total number of fermions, which is a conserved constant of the motion, equal to 1 in these simulations.

Figure~\ref{fig:Excited-TwoStates} shows the dynamics
starting from a superposition of two excited eigenstates.
It shows perfect sinusoidal oscillation over time.
The period of the oscillation is determined from the energy difference of the two eigenstates.
To get maximum density oscillation between sites in this two-site system,
one of the eigenstates should be from total momentum (electron plus phonon) zero
and the other one from $\pi$.
One eigenstate has odd parity, and the other even parity.

\section{Semiclassical}
In the semiclassical approximation, on a periodic 1D chain, the particle will travel through the entire lattice and back to its origin.
In higher dimensions, the particle is no long required to visit every site.
In any finite system in any dimension, the particle will form and continue to repeat a closed trajectory.  This also implies that as the trajectory develops, it will close at its starting point and not some intermediate point, because the trajectory also repeats itself in the distant past.  The trajectory is a loop, not a loop with a tail.  The trajectory can self-intersect at a point, but once it traverses a bond in the same direction for the second time, it must duplicate the already established path.
Here, we focus on how many sites the particle visits
in a large system, and how much time it takes to close the trajectory.

\subsection{Chain of polygons}

\begin{figure}[b]
  \begin{center}
    \includegraphics[width=\columnwidth]{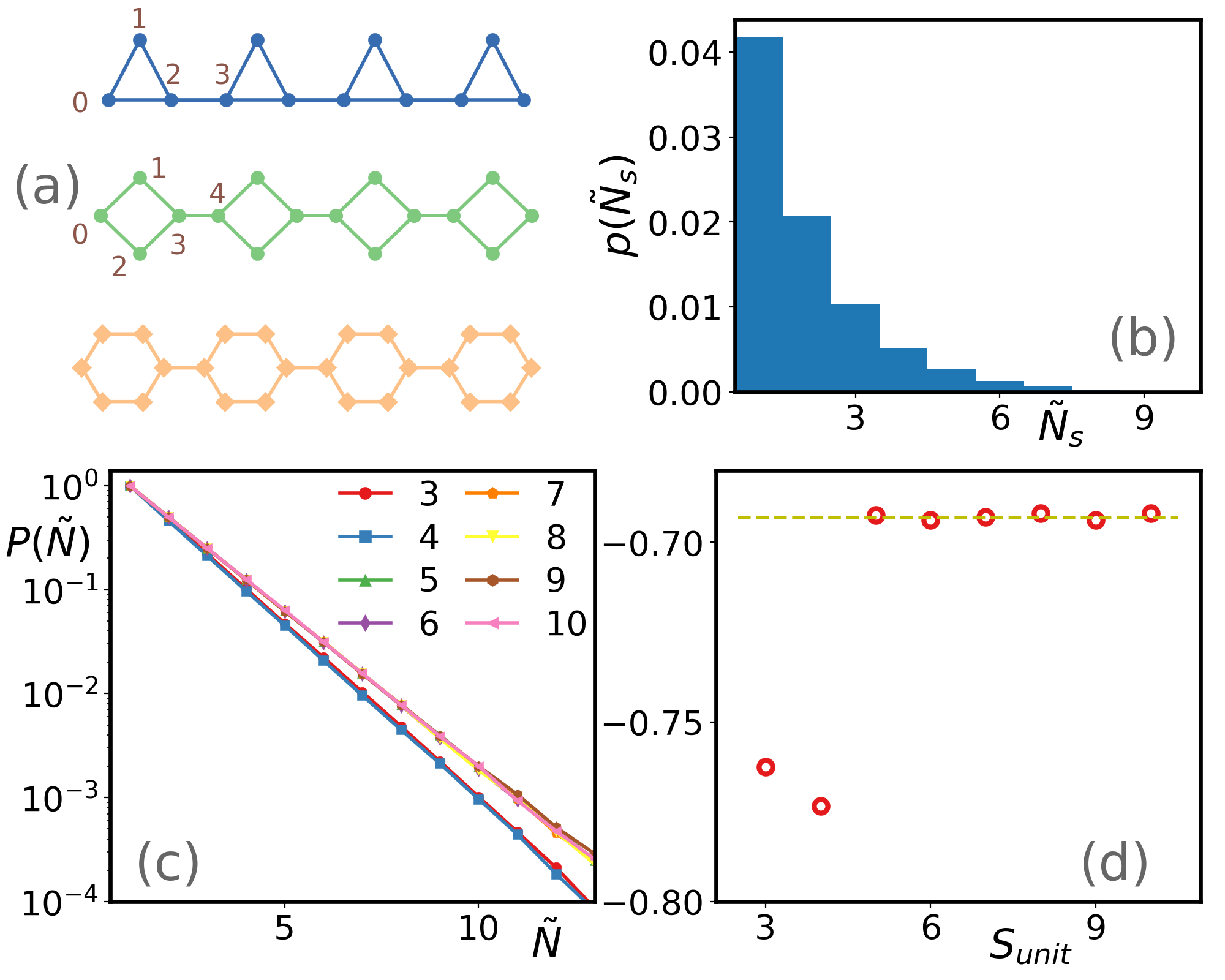}
    \caption{
      (a) Chain of polygons with $S_\text{unit}\!=\!3$-site, $4$-site and $6$-site unit cell.
      (b) Probability of visiting $\tilde{N}_s$ unit cells to close trajectory in a system with $100$ unit cells and $S_\text{unit}\!=\!10$.
      (c) Probability of visiting more than $\tilde{N}$ unit cells to form a closed trajectory, which is the integral of (b), of $100$ unit cells with different $S_\text{unit}$.
      (d) The slope for each $S_\text{unit}$.  The dashed line marks $\!-\!\ln2$.
      Open boundary conditions are used with the particle starting at left edge.
      There are more than $10^5$ realizations for each $S_\text{unit}$.
      The phase is taken as a random variable between $0$ and $2\pi$ and the amplitude is a thermal distribution.
    }
    \label{fig:CPolygons}
  \end{center}
\end{figure}

%
Here, we consider a slightly complicated lattice, a chain of polygons, as illustrated in Fig.~\ref{fig:CPolygons}(a).
This system is still a 1D chain, with a unit cell having more than 2 sites.
In the semiclassical model, the particle will never immediately jump back to the site it came from if there is any alternative.
In this case, it is not necessary for the particle to visit all of the sites before closing the trajectory and repeating itself in the future.
In this example, the possible number of jumps to close a trajectory is discrete $N_s\!=\!\tilde{N}S_\text{unit}+2(\tilde{N}-1)$ where $\tilde{N}$ is the number of unit cells that the particle visits and $S_\text{unit}$ is the number of sites in single unit cell.
Assume the chain has open boundary conditions and the particle starts from the left edge.
The amplitude and phase of the potential on each site is an independent identically distributed random variable.
At each step, the random variables on the neighboring sites are examined to see where the electron moves next.
Considering a chain of hexagons, from the left edge of the hexagon, the electron chooses either the top or the bottom path, each with probability 1/2.
When it reaches the right edge of the hexagon, the electron either continues to the right or closes the hexagon.
Each probability is also 1/2, because the amplitudes and phases on the two new sites are identically distributed.

\begin{equation}\label{eq:PredictPoly}
  p(\tilde{N}_s+1)\!=\!\frac{1}{2}p(\tilde{N}_s).
\end{equation}

This is not, however, true for small polygons.
Consider the square in Fig.~\ref{fig:CPolygons}(a), and assume the electron has taken the path 0-2-3.
The next jump is either to site 1 or site 4.  The amplitude and phase on site 1 were previously examined when the electron was choosing which path to take from site 0.
Site 1 was not chosen because it had an avoided crossing later than that of site 2.
The amplitude and phase on site 1 conditioned on the fact that it was the path not previously chosen now has a different distribution than the standard distribution on site 4.
The result is that the probability to move on to a new polygon is not equal to 1/2 in this case.
The same argument applies to polygons with less than 6 sides, as shown in Fig.~\ref{fig:CPolygons}(b,c,d).
In fact, the same argument applies even to larger polygons, if the leads are attached asymmetrically so there are fewer than 2 vertices between leads (not shown).
We define the integral
\begin{equation}
  P(\tilde{N})=\int_{\tilde{N}}^\infty d\tilde{N}^\prime p(\tilde{N}^\prime),
\end{equation}
which is the probability that the particle visits more than $\tilde{N}$ unit cell(s) to close the trajectory, plotted in Fig.~\ref{fig:CPolygons}(c).
Here, both quantities, $p(\tilde{N}_s)$ and $P(\tilde{N})$, decay exponentially.
Figure~\ref{fig:CPolygons}(c) shows the scaling of $P(\tilde{N})$ with different $S_\text{unit}$'s.
Large unit cell sizes give the same scaling up to statistical error.
More than $10^5$ realizations (different amplitudes and phases) are used for each lattice type.
For unit cell $S_\text{unit}\!\geq\!6$, the slope is $\!-\!\ln2$, shown in Fig.~\ref{fig:CPolygons}(d), consistent with Eq.~\eqref{eq:PredictPoly}.
For $S_{\text{unit}}\!=\!3$, $4$ and $5$, the scaling is different because of the {\it memory} from previous decision making.
